\documentclass[12pt]{article}
\usepackage{amsmath}
\usepackage{mathtools}
\usepackage{graphicx}
\PassOptionsToPackage{hyphens}{url}\usepackage[pdftex]{hyperref}

\RequirePackage{xspace}

\usepackage{relsize}
\def\babar{\mbox{\slshape B\kern-0.1em{\smaller A}\kern-0.1em
    B\kern-0.1em{\smaller A\kern-0.2em R}}}

\def\textcolor#1#2{#2}

\newcommand\pubnumber{DPF2013-125}
\newcommand\pubdate{\today}

\def\caltech{Physics Department 356-48\\
Caltech, Pasadena, CA 91125}

\def\Title#1{\begin{center} {\Large #1 } \end{center}}
\def\Author#1{\begin{center}{ \sc #1} \end{center}}
\def\Address#1{\begin{center}{ \it #1} \end{center}}

\newcommand\pubblock{\rightline{\begin{tabular}{l} \pubnumber\\
         \pubdate  \end{tabular}}}
\newenvironment{Abstract}{\begin{quotation}  }{\end{quotation}}
\newenvironment{Presented}{\begin{quotation} \begin{center} 
             PRESENTED AT\end{center}\bigskip 
      \begin{center}\begin{large}}{\end{large}\end{center} \end{quotation}}
\def\Acknowledgments{\bigskip  \bigskip \begin{center} \begin{large}
             \bf ACKNOWLEDGMENTS \end{large}\end{center}}

\def\beq{\begin{equation}}
\def\eeq#1{\label{#1}\end{equation}}
\def\eeqn{\end{equation}}

\def\beqa{\begin{eqnarray}}
\def\eeqa#1{\label{#1}\end{eqnarray}}
\def\eeqan{\end{eqnarray}}

\let\bar=\overbar

\def\Dslash{\not{\hbox{\kern-4pt $D$}}}
\def\dslash{\not{\hbox{\kern-2pt $\del$}}}

\def\msb{{\bar{\ssstyle M \kern -1pt S}}}

\begin{document}
\begin{titlepage}
\pubblock

\vfill
\Title{Hadron production in $e^+e^-$ annihilation at \babar, and implication for the muon anomalous magnetic moment}
\vfill
\Author{Frank C. Porter\\
For the \babar\ Collaboration}
\Address{\caltech}
\vfill
\begin{Abstract}
The \babar\ collaboration has an extensive program of studying hadronic
cross sections in low-energy $e^+e^-$ collisions, accessible via
initial-state radiation. Our measurements allow significant improvements
in the precision of the predicted value of the muon anomalous magnetic
moment. These improvements are necessary for illuminating the
current ~3.6 sigma difference between the predicted and the experimental
values. We have published results on a number of processes with two to
six hadrons in the final state. We report here the results of recent studies
with final states that constitute the main contribution to 
the hadronic cross section in the energy region between 1 and 3 GeV,
 as $e^+e^- \to K^+K^-$,  $\pi^+\pi^-$,  and $e^+e^- \to 4 \hbox{ hadrons}$.

\end{Abstract}
\vfill
\begin{Presented}
DPF 2013\\
The Meeting of the American Physical Society\\
Division of Particles and Fields\\
Santa Cruz, California, August 13--17, 2013\\
\end{Presented}
\vfill
\end{titlepage}
\def\thefootnote{\fnsymbol{footnote}}
\setcounter{footnote}{0}

\section{Introduction}

\babar\ is a high luminosity ($\sim10^{34}\hbox{ cm}^{-2}\hbox{s}^{-1}$) $e^+e^-$ experiment at the PEP-II asymmetric storage ring located at SLAC.
In processes involving initial state radiation, this enables precise measurement of 
$\sigma(e^+e^-\to\hbox{ hadrons})$ as a function of CM energy from threshhold to several GeV. 
These measurments provide the opportunity for precise determination of hadronic form factors, in particular for $\pi$, $K$, and $p$, and for studies of light hadron spectroscopy.
Here, we emphasize the important role these measurements have as inputs to the standard model (SM) calculation of the hadronic vacuum polarization (HVP) contribution
to the muon anomalous magnetic moment, $(g-2)_\mu$.

The magnetic moment of a lepton, $\ell$, of mass $m_\ell$ and charge $e$ may be written in the form
\begin{equation}
\vec{\mu}_\ell = - \frac{g_\ell e }{ 2 m_\ell}\vec{S},
\end{equation}
where $\vec{S}$ is the spin angular momentum of the lepton. The ``$g$-factor'', $g_\ell$, is
predicted to be two according to the Dirac equation, but higher order corrections yield
deviations. These deviations are expressed in the magnetic moment anomaly,
\begin{equation}
a_\ell\equiv \frac{(g_\ell - 2)}{2}.
\end{equation}

Interest in $a_\ell$ centers around its sensitivity to possible new physics (NP). As a helicity-flip
process, the sensitivity to NP depends on lepton mass as $\sim m_\ell^2$. In spite of the very precise
measurement of $a_e$, the $m_\ell^2$ factor wins, and the muon anomaly is presently more sensitive in these terms. The $\tau$ is still heavier, but is short-lived and precise measurement
of $a_\tau$ is currently impractical.

The currently most precise measurement of the muon anomaly and its comparison with the SM prediction are~\cite{bib:Bennett2006,bib:Engel2012,bib:DHMZ}:
\begin{align}
a_\mu(\hbox{measured}) &= 116592089\pm 63\times 10^{-11},\\
a_\mu(\hbox{SM}) &= 116591802\pm 49\times 10^{-11}.
\end{align}
Thus, the measured value is 3.6$\sigma$ larger than the SM prediction, and deserves
investigation.

The standard model prediction has several important components~(e.g., \cite{bib:Engel2012,bib:DHMZ} and references therein):
\begin{align}
a_\mu(\hbox{SM}) &= a_\mu(\hbox{QED}) + a_\mu(\hbox{weak}) + a_\mu(\hbox{had}),\\
a_\mu(\hbox{QED}) &=116584718.10\pm0.15\times 10^{-11},\\
a_\mu(\hbox{weak}) &=154\pm2\times 10^{-11},\\
a_\mu(\hbox{had}) &=6930\pm49\times 10^{-11}.
\end{align}
The hadronic (``had'') component is the largest component after $a_\mu(\hbox{QED})$, and 
is by far the dominant source of uncertainty in the SM prediction. This component in turn
has two contributions, from hadronic vacuum polarization (HVP) and hadronic light-by-light scattering.
The uncertainties from these two components are of the same order, but the largest uncertainty ($\pm 42\times 10^{-11}$) is from the hadronic vacuum polarization, $a_\mu(\hbox{HVP})$.
It is not possible to compute $a_\mu(\hbox{HVP})$ perturbatively.
Instead, we may measure $\sigma(e^+e^-\to\hbox{hadrons})$ as a function of CM energy and use dispersion relations to extract $a_\mu(\hbox{HVP})$.

\begin{figure}[htb]
\centering
\includegraphics[height=1.in]{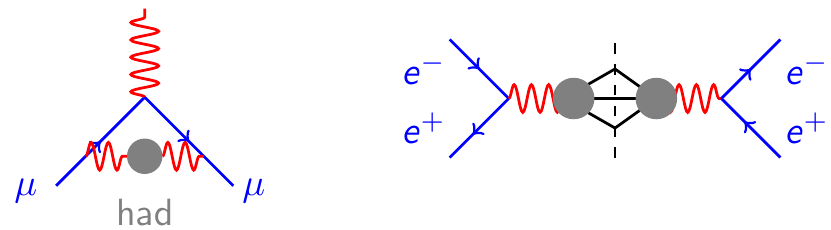}
\caption{Relating $a_\mu(\hbox{HVP})$ to $e^+e^-\to\hbox{ hadrons}$ scattering via dispersion relations. The blobs represent hadronic systems.}
\label{fig:dispDiag}
\end{figure}

The dispersion relation for $a_\mu(\hbox{had})$ may be written:
\begin{equation}
a_\mu(\hbox{had}) = \frac{\alpha^2}{3\pi^2}\int_{\hbox{threshold}}^\infty R(s) \frac{K(s)}{s} ds
\end{equation}
where 
\begin{equation}
R(s) = \frac{\sigma^0(e^+e^-\to\hbox{hadrons}(\gamma))}{\sigma_{\rm pt}}
\end{equation}
and~\cite{bib:Davier}
\begin{equation}
K(s)\sim m_\mu^2/3s.
\end{equation}
The quantity $\sigma^0$ is the bare cross section, excluding vacuum polarization effects, but including final state radiation (FSR).
The idea behind the approach is seen in Fig.~\ref{fig:dispDiag}. Because of the $\sim 1/s^2$ weighting on $R$, the emphasis is from the low-energy portion of the hadron spectrum.
Hence, the
dominant contribution is from $\pi^+\pi^-$. However, other channels cannot be neglected
at the required precision.

\section{The ISR method}

\begin{figure}[htb]
\centering
\includegraphics[height=0.85in]{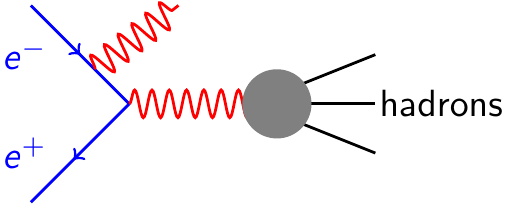}\hskip15pt
\includegraphics[height=0.85in]{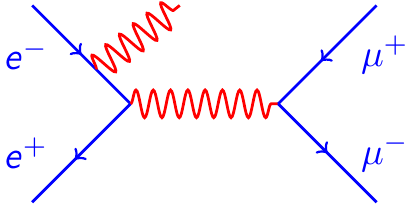}
\caption{Use of ISR in $e^+e^-$ scattering to measure $\sigma(e^+e^-\to\hbox{ hadrons})$ (left) and $\sigma(e^+e^-\to\mu^+\mu^-)$ (right) as a function of the invariant mass of the virtual photon.}
\label{fig:ISRdiag}
\end{figure}

To implement this approach, we need to measure $\sigma^0$ as a function of $s$. We may achieve this in a single
$e^+e^-$ experiment by making use of initial state radiation (ISR). The idea is illustrated in 
Fig.~\ref{fig:ISRdiag}.

Most of the \babar\ data is for  $e^+e^-$ collisions at $\sqrt{s}=10.6$ GeV.
With ISR, the effective $e^+e^-\to\gamma^*$ energy is $\sqrt{s^\prime} = \sqrt{s(1-x)}$, where $x=2E^*_\gamma/\sqrt{s}$ in the CM frame. 
Events are selected with a high energy  ISR photon ($E^*_\gamma > 3$ GeV) at large angle.  
The ISR photon is opposite the hadrons in the CM. Thus, there is high acceptance for boosted hadrons even from threshhold.
Additional ISR and FSR must be accounted for.
This technique provides a measurement from threshold to 3--5 GeV in a single dataset, and reduces systematics.
\babar\ has an extensive program to measure $e^+e^-\to\hbox{hadrons}$ as a function of energy using this ISR method, as shown in Table~\ref{tab:BaBarISR}
(channels include a possible additional FSR photon).

\begin{table}[ht]
\begin{center}
\begin{tabular}{ll}
\hline
\textcolor{blue}{Final state(s)} & \textcolor{blue}{Publication}\\
\hline
\small{\textcolor{orange}{$\pi^+\pi^-$}} & {\textcolor{orange}{PRD {\bf 86} 032013 (2012)}}\\ 
\textcolor{red}{$K^+K^-$} & PRD {\bf 88} 032013 (2013)\\ 
$\pi^+\pi^-\pi^0$ & PRD {\bf 70}  072004 (2004)\\ 
$K^+K^-\eta$, $K^+K^-\pi^0$, $K^0_SK^\pm\pi^\mp$ & PRD {\bf 77} 092002 (2008)\\
\textcolor{red}{$\pi^+\pi^-\pi^+\pi^-$} & \textcolor{red}{PRD {\bf 85} 112009 (2012)}\\ 
\textcolor{red}{$K^+K^-\pi^+\pi^-$, $K^+K^-\pi^0\pi^0$, $2(K^+K^-)$} & \textcolor{red}{PRD {\bf 86} 012008 (2012)}\\ 
\small{$\Lambda\bar\Lambda$, $\Lambda\bar\Sigma^0$, $\Sigma\bar\Sigma^0$} & {PRD {\bf 76} 092006 (2007)}\\
$2(\pi^+\pi^-)\pi^0$, $2(\pi^+\pi^-)\eta$, $K^+K^-\pi^+\pi^-\pi^0$, &\\ \qquad $K^+K^-\pi^+\pi^-\eta$ & PRD {\bf 76} 0922005 (2007)\\ 
$\phi\eta$ & PRD RC {\bf 74} 111103 (2006)\\
$3(\pi^+\pi^-)$, $2(\pi^+\pi^-\pi^0)$, $K^+K^-2(\pi^+\pi^-)$ & PRD {\bf 73} 052003 (2006)\\ 
\textcolor{red}{$p\bar p$ (C.\ Cartaro, these proceedings)} & PRD {\bf 87} 092005 (\textcolor{red}{2013})\\
\textcolor{magenta}{$K^0_SK^0_L$, $K^0_SK^0_L\pi^+\pi^-$, $K^0_SK^\pm\pi^\mp\pi^0$,}& \\   \qquad\textcolor{magenta}{$K^0_SK^\pm\pi^\mp\eta$, $\pi^+\pi^-2\pi^0$} & \textcolor{magenta}{in progress}\\
\hline
\end{tabular}
\caption{\babar\ ISR measurements of $e^+e^-\to\hbox{hadrons}$.}
\label{tab:BaBarISR}
\end{center}
\end{table}

As an example of the analysis strategy, we consider the recently published $K^+K^-(\gamma)$ channel~\cite{bib:LeesKK}. The $K^+K^-(\gamma)$ yield is measured in ISR production.
The effective luminosity is obtained from the simultaneously measured $\mu^+\mu^-(\gamma)$ rate. This approach is used for the two-prong $\pi^+\pi^-(\gamma)$ channel as well.
The efficiency is estimated  with data-corrected simulations.
Equation~\ref{eq:measStrategy} gives the relation from which the cross section is determined.
\begin{equation}
\textcolor{blue}{\frac{dN_{K^+K^-(\gamma)\gamma_{\rm ISR}}}{d\sqrt{s^\prime}}} =
 \textcolor{red}{\frac{dL^{\rm eff}_{\rm ISR}}{d\sqrt{s^\prime}}}\ 
 \textcolor{orange}{\epsilon_{KK\gamma_{\rm ISR}}(\sqrt{s^\prime})}\
 \textcolor{magenta}{\sigma^0_{KK(\gamma)}(\sqrt{s^\prime})}
\label{eq:measStrategy}
\end{equation}
The ``bare'' cross section $\sigma^0$ includes final state radiation (FSR), but no
leptonic or hadronic vacuum polarization effects. These have been removed by
using the normalization based on the measured $\mu^+\mu^-(\gamma)$ rate.

The systematic uncertainties in efficiency and background estimation must be 
carefully controlled to avoid exceeding the available statistical precision. The interested
reader is referred to the primary publications for details; we only provide a summary here.
The MC efficiency is corrected for MC/data differences, using in situ
efficiency measurements. The corrections are in four categories, with associated $s^\prime $-dependent systematic uncertainties:
({\it i\/}) Trigger corrections are of order $\sim \hbox{few}\times 10^{-4}$, contributing small systematic uncertainty; 
({\it ii\/}) Corrections for tracking result in systematic uncertainties $<$ few $\times 10^{-3}$;
({\it iii\/}) Particle identification corrections result in systematic uncertainties typically a few $\times 10^{-3}$
({\it iv\/}) Kinematic fit selection uncertainties result from possible errors in the modeling of additional ISR/FSR: $<$ few $\times 10^{-3}$. 

Backgrounds arise mainly from cross-feed from other ISR processes. The systematic uncertainty in the background subtraction is typically a
few $\times 10^{-3}$ or less depending on channel, but tends to be higher at extremes of
$\sqrt{s^\prime}$.

\section{Results}

The $K^+K^-$ and $\pi^+\pi^-$ results are based on the first 232 fb$^{-1}$ of \babar\ data; the other results below use a 454 fb$^{-1}$ dataset. The luminosity normalization for both the 
$K^+K^-$ and $\pi^+\pi^-$ is taken from the simultaneous $\mu^+\mu^-$ measurment.
For the other channels, the standard \babar\ luminosity determination,~\cite{bib:lumiNIM}, is used. In this case, the result is the dressed cross section, including vacuum polarization, which must be 
corrected for in computing $a_\mu$.

\subsection{$e^+e^-\to K^+K^-(\gamma)$}

\begin{figure}[htb]
\centering
\includegraphics [width=0.49\textwidth]{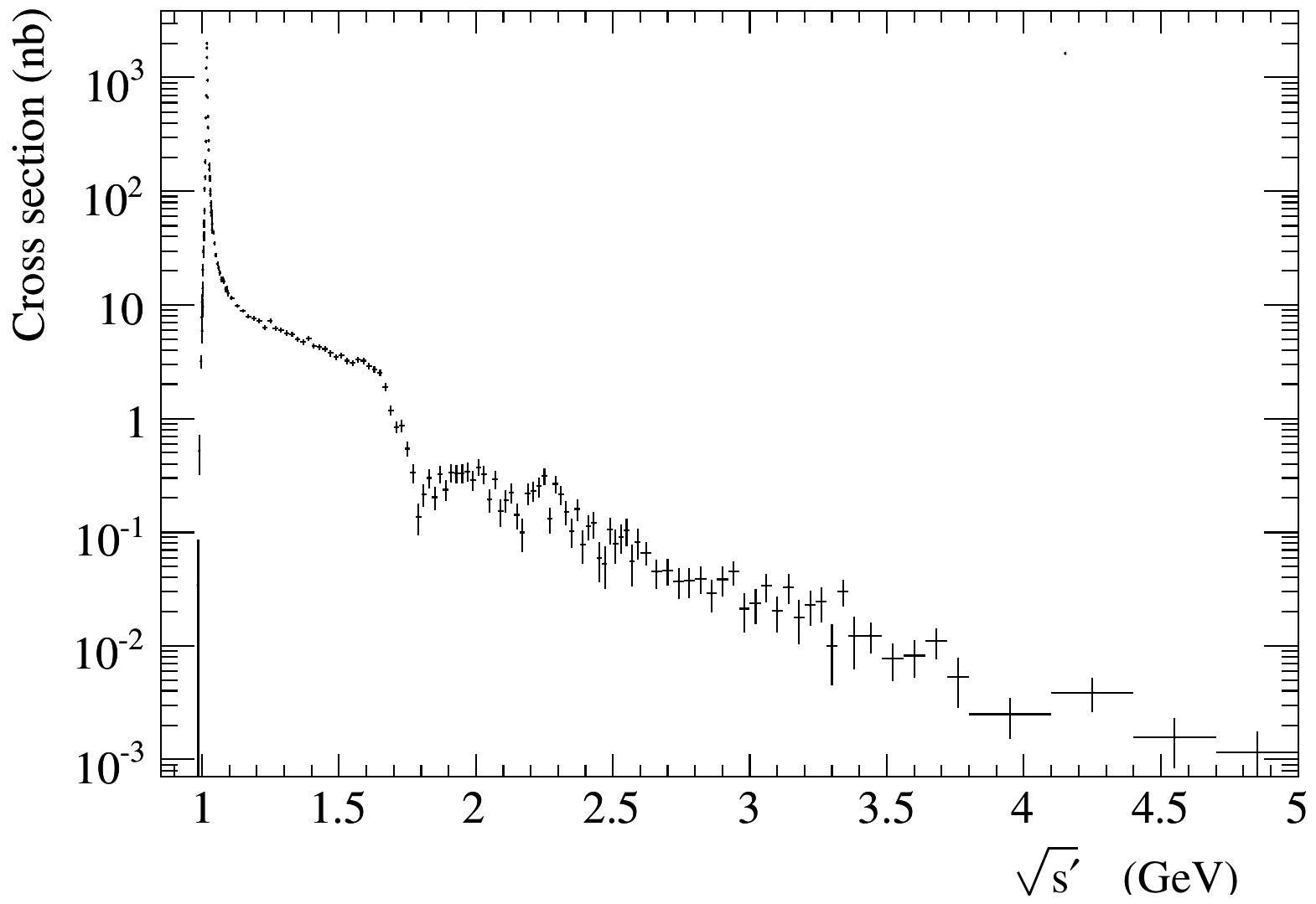}\hskip.1cm
\includegraphics [width=0.49\textwidth]{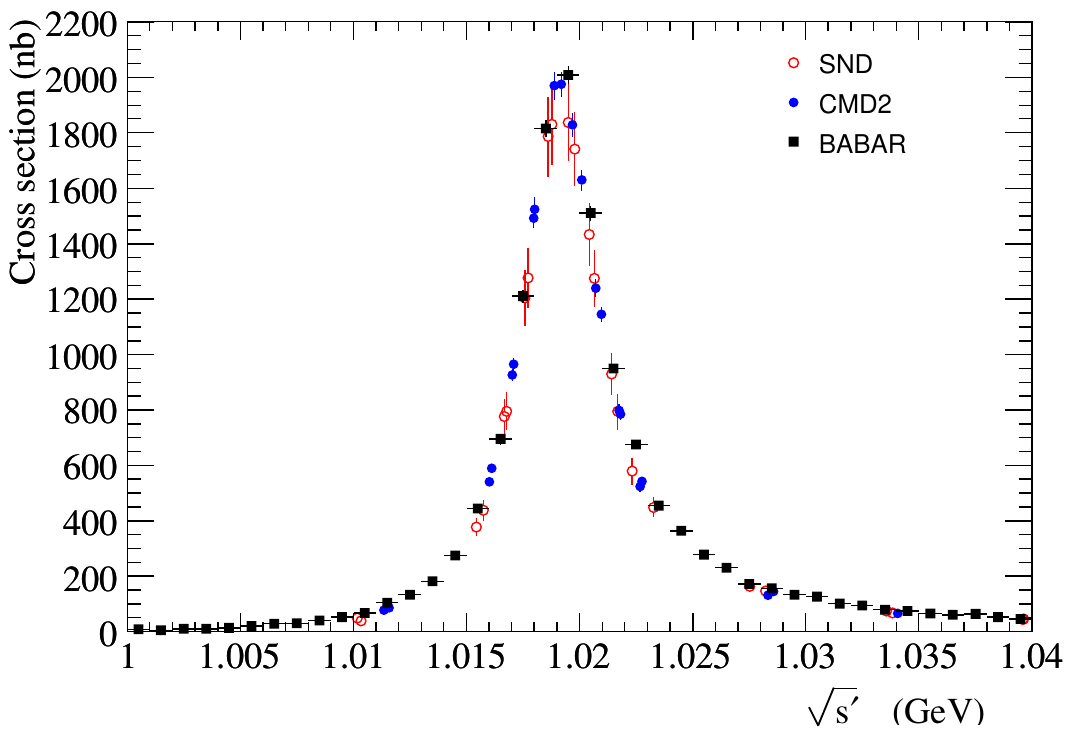}
\caption{The cross section $\sigma^0(e^+e^-\to K^+K^-(\gamma))$ as a function of $\sqrt{s^\prime}$. Left:
\babar\ result from threshold to 5 GeV. Right: Comparison of \babar\ result with previous results in the $\phi$ region. From~\cite{bib:LeesKK}.}
\label{fig:KKcs}
\end{figure}

\begin{figure}[htb]
\centering
\includegraphics [width=.99\textwidth]{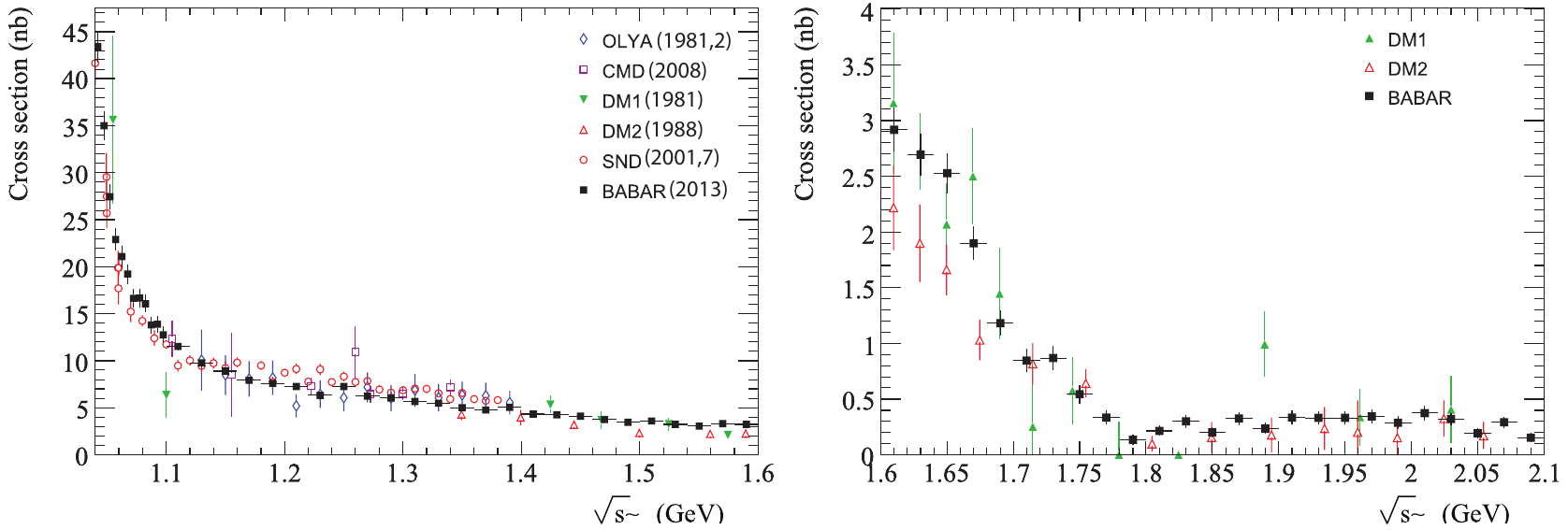}
\caption{Comparison of the \babar\ $e^+e^-\to K^+K^-(\gamma)$ result with previous experiments~\cite{bib:LeesKK}.}
\label{fig:KKcsComp}
\end{figure}

The bare cross section (including FSR) for $K^+K^-(\gamma)$ is shown in Figs.~\ref{fig:KKcs} and~\ref{fig:KKcsComp}, including comparison with earlier results. Here, the $J/\psi$ and $\psi(2S)$ have been subtracted, as these are treated separately. While similar with the earlier measurements, there are significant differences in normalization at the $\phi$ resonance, and in the comparison with SND and DM2 at higher $\sqrt{s^\prime}$.

\begin{figure}[htb]
\centering
\includegraphics [width=0.5\textwidth]{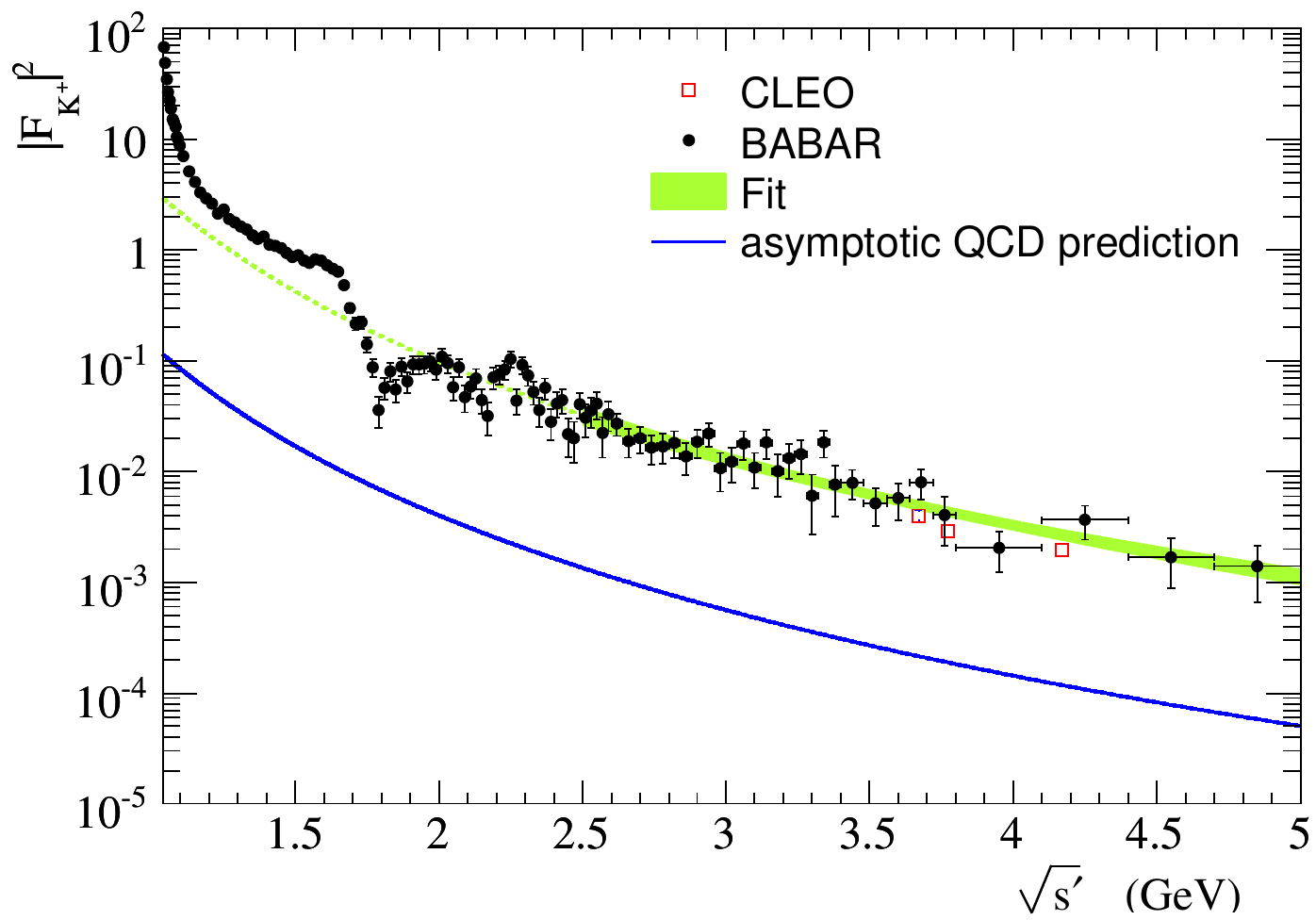}
\caption{The square of the charged kaon form factor vs $\sqrt{s^\prime}$, including comparison with CLEO and asymptotic QCD~\cite{bib:LeesKK}.}
\label{fig:KchFF}
\end{figure}

Figure \ref{fig:KchFF} shows the result for the charged kaon form factor, which is consistent in the
3--4 GeV region with earlier results from CLEO. Asymptotic QCD predicts an $s^\prime$ dependence of~\cite{bib:QCDKFF}
\begin{equation}
F_K(s^\prime) = 16\pi \alpha_s(s^\prime)\frac{f_K^2}{s^\prime}.
\end{equation}
This prediction (blue curve) is shown in the figure; the prediction for $\left| F_K\right|$ falls about a factor of four below the data.
The shape is however consistent with with predicted $|F_K|^2\propto {s^\prime}^{-2}$ fall-off (power law fit at high $s^\prime$ shown by the green band). The discrepancy in normalization is presently not 
well-understood.

\subsection{$\pi^+\pi^-(\gamma)$ cross section results}

\begin{figure}[htb]
\centering
\includegraphics [width=0.45\textwidth]{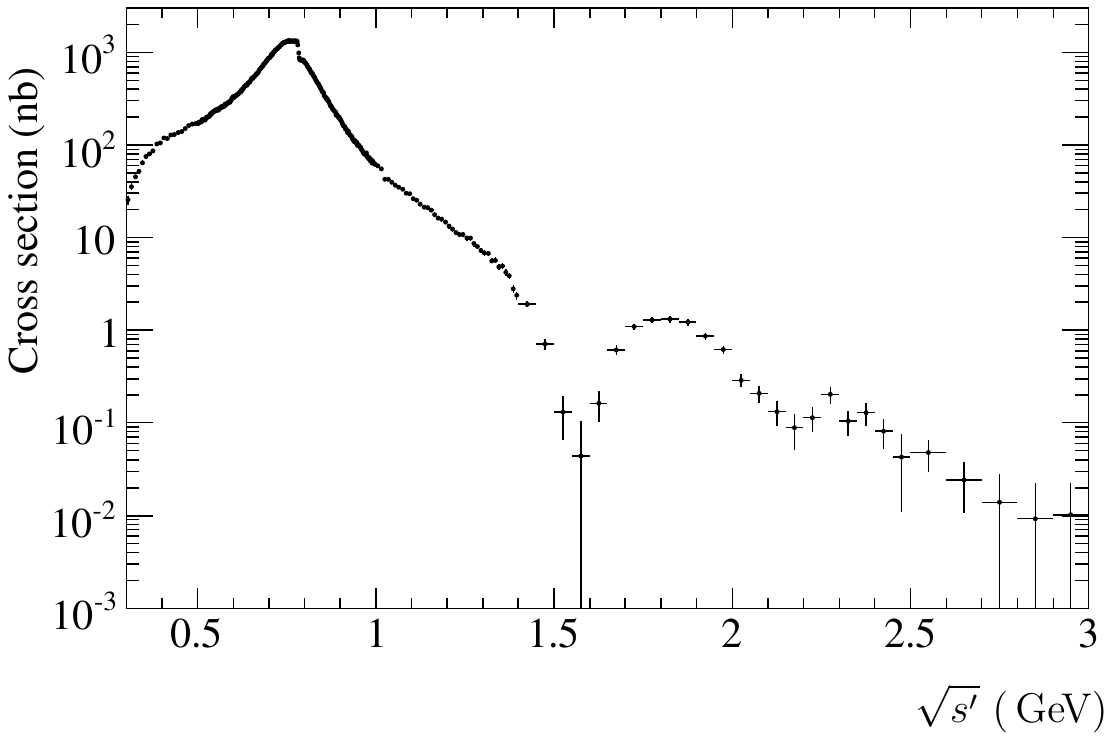}\hskip10pt
\includegraphics [width=0.45\textwidth]{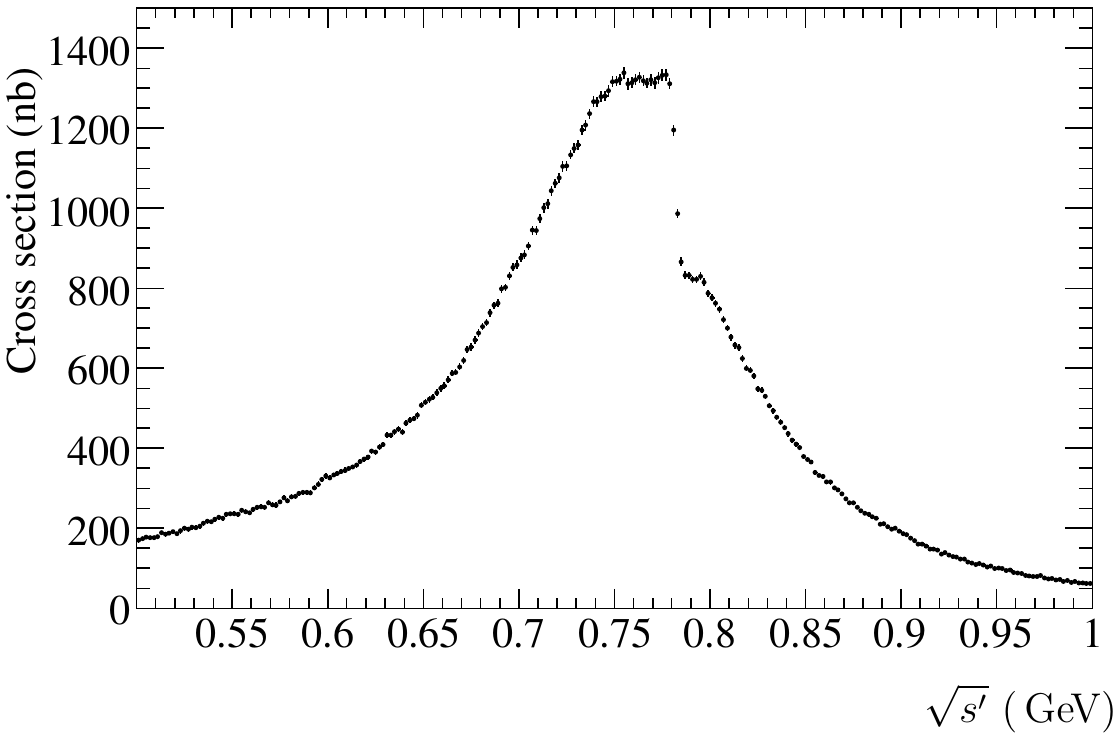}
\caption{Left: The \babar\ result for the bare $e^+e^-\to \pi^+\pi^-(\gamma)$ cross section vs $\sqrt{s^\prime}$. Right: The \babar\ result for the bare $e^+e^-\to \pi^+\pi^-(\gamma)$ cross section vs $\sqrt{s^\prime}$ in the $\rho/\omega$ region. From~\cite{bib:Leespipi}.}
\label{fig:pipi}
\end{figure}

The analysis of the dominant $\pi^+\pi^-(\gamma)$ channel is very similar with that for the $K^+K^-$ channel. The bare cross section (including FSR) is shown in Fig.~\ref{fig:pipi}.

\subsection{$K^+K^-\pi\pi$ cross section results}

\begin{figure}[htb]
\centering
\includegraphics [width=0.38\textwidth]{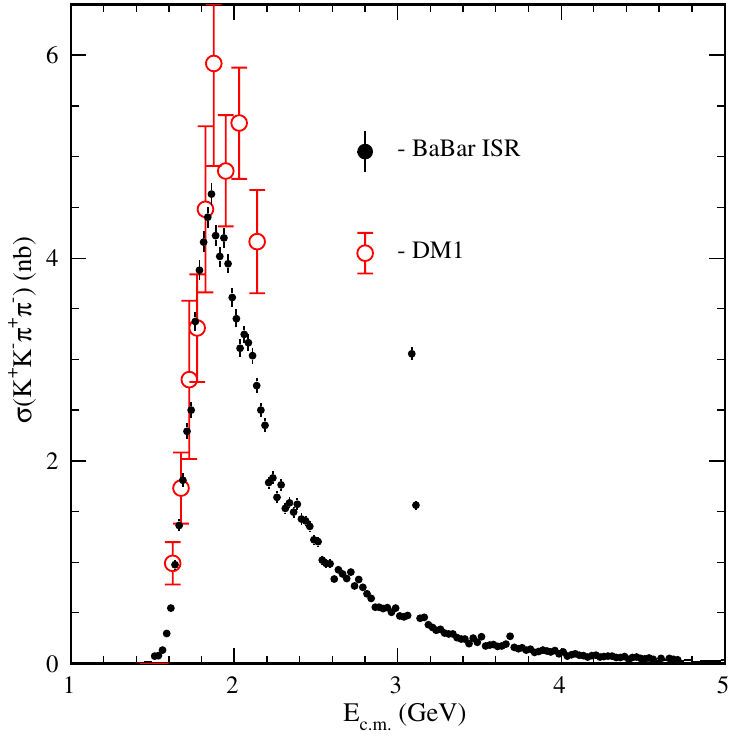}\hskip15pt
\includegraphics [width=0.4\textwidth]{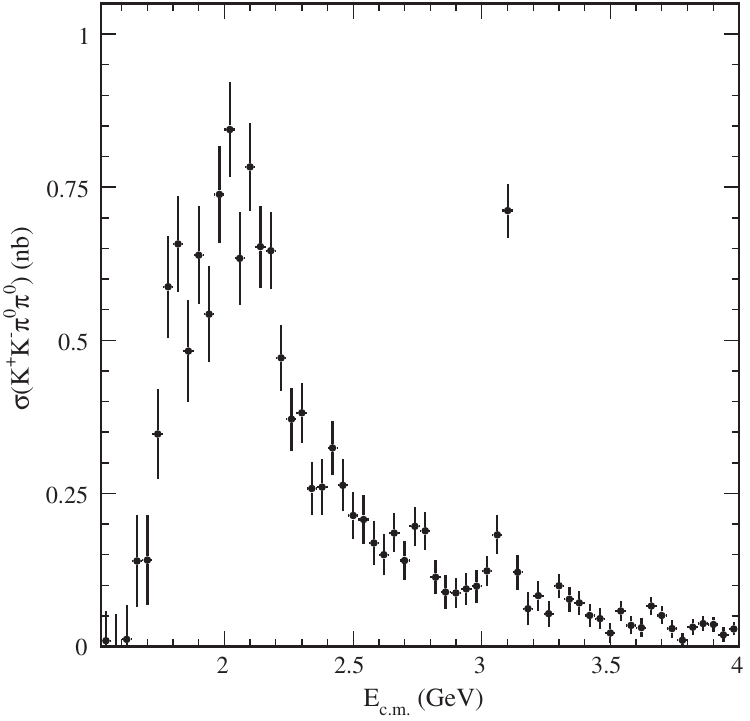}
\caption{Left: $\sigma(K^+K^-\pi^+\pi^-)$. Right: $\sigma(K^+K^-\pi^0\pi^0)$. Errors shown are statistical only. From~\cite{bib:KKpipi}.}
\label{fig:KKpipics}
\end{figure}

Based on a 454 fb$^{-1}$ dataset, the
dressed cross section measurements from \babar\ for $e^+e^-\to K^+K^-\pi\pi$ are shown in Fig.~\ref{fig:KKpipics} (statistical uncertainties shown).
The $K^+K^-K^+K^-$ channel has also been measured, but is not shown here.
The cross section at high $s^\prime$ for $K^+K^-\pi^+\pi^-$ is systematically smaller than the earlier DM1 result.

\subsection{$\pi^+\pi^-\pi^+\pi^-$ cross section results}

\begin{figure}[htb]
\centering
\includegraphics[width=.4\textwidth]{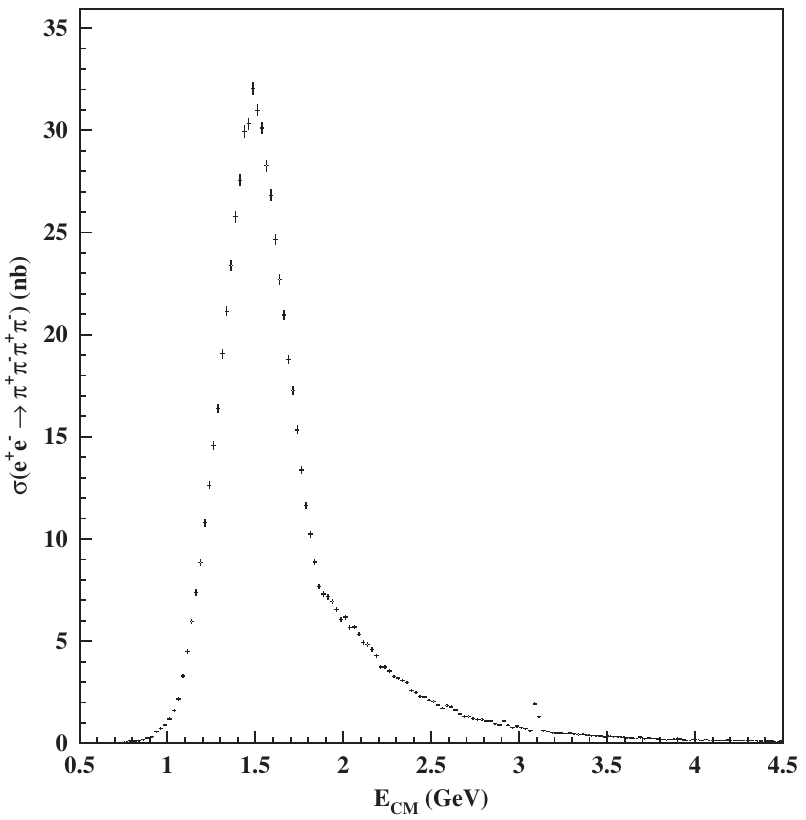}\hskip15pt
\includegraphics[width=.41\textwidth]{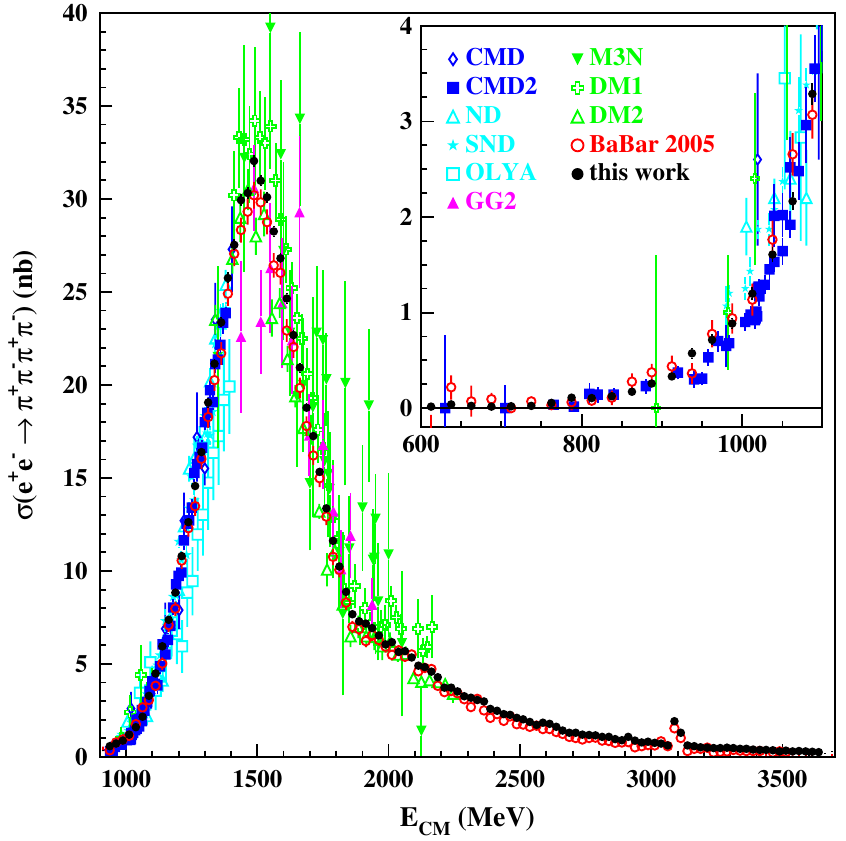}
\caption{Results for the $e^+e^-\to \pi^+\pi^-\pi^+\pi^-$ cross section. Left: \babar\ results. Right: comparison with earlier results.  Errors shown are statistical only. From~\cite{bib:4pi}.}
\label{fig:4pics}
\end{figure}

Based on a 454 fb$^{-1}$ dataset, the
dressed cross section from \babar\ for $e^+e^-\to \pi^+\pi^-\pi^+\pi^-$ is shown in Fig.~\ref{fig:4pics} (statistical uncertainties shown). Our results are consistent with but more precise than the previous results.

\section{Discussion}

Three of the dominant contributions to $a_\mu(\hbox{HVP})$, with cross section measurements reported here, are shown in Table~\ref{tab:summary}.
The \babar\ precision for  $\pi^+\pi^-$ is comparable with the previous world average,
for 4$\pi$ it is a factor of 2.6 better, and for $K^+K^-$ it is a factor of 3 better.

\begin{table}[ht]
\begin{center}
\begin{tabular}{|l|c|c|}
\hline
Channel & \multispan{2}{\hfil$a_\mu(\hbox{HVP})\ (10^{-11})$}\hfil\vline \\
& \babar & world average w/o \babar\\
\hline
$\pi^+\pi^-$ & $5141\pm22\pm31$ & $5056\pm30$~\cite{bib:DHMYZ}\\
$\pi^+\pi^-\pi^+\pi^-$ & $136.4\pm0.3\pm3.6$ & $139.5\pm9.0\pm2.3$~\cite{bib:DEHZ}\\
$K^+K^-$ & $229.3\pm1.8\pm2.2$ & $216.3\pm2.7\pm6.8$~\cite{bib:DHMZ}\\
\hline
\end{tabular}
\caption{\babar\ results for $a_\mu(\hbox{HVP})$, and comparison with the world averages
excluding \babar.}
\label{tab:summary}
\end{center}
\end{table}

In order to make progress on the experimental measurement, a new experiment, FNAL E989~\cite{bib:E989},
is currently under construction, using upgraded components from the BNL experiment. The
goal of the new experiment is reduce the  uncertainty on the measured $a_\mu$ from $63\times10^{-11}$ to $16\times10^{-11}$.

It is desirable to match this experimental improvement with corresponding improvement
in the precision of the SM prediction. We expect lattice calculations to eventually provide precise SM predictions for HVP. However, on the time scale of E989 the anticipated improvements
in lattice calculations will lead to uncetainties of a ``few percent''~\cite{bib:VdW, bib:ProjectXWP}, which 
is not sufficiently precise.
The present uncertainty on HVP from $e^+e^-$ measurements is already less than a percent.
Matching the projected experimental precision of $16\times10^{-11}$ requires HVP to be computed to $\sim0.2$\%. It will be difficult to achieve this even with $e^+e^-$ in the desired time frame.
However, it may be possible to make progress with data already in hand. The
dominant $\pi\pi$ channel result is on half of the \babar\ dataset. 
It may be possible to use the other
half as well on the E989 timescale, perhaps with gains in both statistical and systematic
precision.

\Acknowledgments
I am grateful to my \babar\ colleagues for many stimulating discussions.
This work is supported in part by the U.~S.\ Department of Energy under
grant DE-FG02-92-ER40701.

\end{document}